\documentclass[authoryear,preprint,1p]{elsarticle}

\usepackage[authoryear]{natbib}
\usepackage{amssymb,amsthm,epsfig}
\usepackage[centertags]{amsmath}
\usepackage{bm}
\newcommand{\Es}{\mathbb{E}}
\newcommand{\diag}{{\rm diag}}

\begin{document}

\begin{frontmatter}
\title{Size and power properties of some tests in the Birnbaum--Saunders regression model}
\author[A]{Artur J.~Lemonte}
\author[A]{Silvia L.P.~Ferrari\corref{cor1}}
\ead{silviaferrari.usp@gmail.com}
\cortext[cor1]{Corresponding author}
\address[A]{Departamento de Estat\'istica, Universidade de S\~ao Paulo, Rua do Mat\~ao, 1010,
S\~ao Paulo/SP, 05508-090, Brazil}
\begin{abstract}

The Birnbaum--Saunders distribution has been used quite
effectively to model times to failure for materials subject
to fatigue and for modeling lifetime data.
In this paper we obtain asymptotic expansions, up to
order $n^{-1/2}$ and under a sequence of Pitman alternatives,
for the nonnull distribution functions
of the likelihood ratio, Wald, score and gradient
test statistics in the Birnbaum--Saunders
regression model. The asymptotic distributions of all four statistics are
obtained for testing a subset of regression
parameters and for testing the shape parameter.
Monte Carlo simulation is presented in order to compare the
finite-sample performance of these tests.
We also present an empirical application.

\end{abstract}
\begin{keyword}
Birnbaum--Saunders distribution\sep Fatigue life distribution\sep Gradient test\sep
Lifetime data\sep Likelihood ratio test\sep Local power\sep Score test\sep Wald test.
\end{keyword}
\end{frontmatter}

\section{Introduction}\label{introduction}

The Birnbaum--Saunders ($\mathcal{BS}$) distribution has received
considerable attention in the last few years. It was proposed by
\cite{BSa1969a, BSa1969b} and is also known as the fatigue life
distribution. It describes the total
time until the damage caused by the development and growth of
a dominant crack reaches a threshold level and causes a failure.
The random variable $T$ is said to have a $\mathcal{BS}$ distribution
with parameters $\alpha$ and $\eta$,
denoted by $T\sim\mathcal{BS}(\alpha,\eta)$,
if its probability density  function is given by
\[
f(t;\alpha,\eta)=\kappa(\alpha,\eta)t^{-3/2}(t+\eta)
\exp\left\{-\frac{\tau(t/\eta)}{2\alpha^2}\right\}, \qquad t>0,
\]
where $\kappa(\alpha,\eta)=\exp(\alpha^{-2})/(2\alpha\sqrt{2\pi\eta})$,
$\tau(z)=z+z^{-1}$, $\alpha>0$ (shape parameter) and $\eta>0$ (scale parameter).
It is positively skewed, the skewness decreasing with $\alpha$.
For any constant $k > 0$, it follows that $kT \sim\mathcal{BS}(\alpha, k\eta)$.
It is also noteworthy that the reciprocal property holds:
$T^{-1}\sim\mathcal{BS}(\alpha, \eta^{-1})$,
which is in the same family of distributions \citep{Saunders1974}.
There are several recent articles considering the $\mathcal{BS}$ distribution;
see for example, \cite{WuWong2004}, \cite{Diaz-Leiva05},
\cite{LCNV07, LSCN08}, \cite{Balakrishnan-et-al-2007},
\cite{kundu-et-al-2008}, \cite{Leiva-et-al-2008}, \cite{GPB09}, \cite{guiraud-et-al-2009},
\cite{Leiva-et-al-2009}, \cite{XiTang10}, \cite{Meintanis2010}, among others.

\cite{RiekNedelman91} introduced a log-linear regression model based on the
$\mathcal{BS}$ distribution by showing that if
$T\sim\mathcal{BS}(\alpha,\eta)$, then $Y = \log(T)$ has a sinh-normal
distribution with shape, location and scale parameters given by $\alpha$,
$\mu=\log(\eta)$ and $\sigma=2$, respectively, say $Y\sim\mathcal{SN}(\alpha,\mu,2)$.
The regression model proposed by the authors is given by
\begin{equation}\label{eq1}
y_{i} = \bm{x}_{i}^\top\bm{\beta} + \varepsilon_{i},\qquad i = 1, \ldots, n,
\end{equation}
where $y_{i}$ is the logarithm of the $i$th observed lifetime,
$\bm{x}_{i}^{\top} = (x_{i1}, \ldots, x_{ip})$ contains the $i$th
observation on $p$ covariates ($p < n$),
$\bm{\beta} = (\beta_{1}, \ldots, \beta_{p})^\top$ is a vector of
unknown regression parameters, and $\varepsilon_{i}\sim\mathcal{SN}(\alpha, 0, 2)$.
Diagnostic tools for the $\mathcal{BS}$
regression model can be found in \cite{Galea-etal-2004}, \cite{XiWei07} and \cite{Leiva-etal-2007}.

In the $\mathcal{BS}$ regression model hypothesis testing inference is usually
performed using the likelihood ratio, Rao score and Wald tests.
A new criterion for testing hypothesis, referred to as the
{\it gradient test}, has been proposed by \cite{Terrell2002}.
Its statistic shares the same first order asymptotic properties with the
likelihood ratio, Wald and score statistics
and is very simple when compared with the other three classic tests. In fact, \cite{Rao2005}
wrote: ``The suggestion by Terrell is attractive as it is simple to compute.
It would be of interest to investigate the performance of the [gradient] statistic.''
To the best of our knowledge, however, there is no mention in the
statistical literature on the use of the gradient test in $\mathcal{BS}$ regressions.

In this paper we compare the four rival tests from two different points of view. First, we
invoke asymptotic arguments. We then move to a finite-sample comparison, which is
accomplished by means of a simulation study. Our principal aim is to help practitioners to
choose among the different criteria when performing inference in $\mathcal{BS}$ regressions.

On asymptotic grounds, it is known that, to the first order of approximation, the likelihood ratio,
Wald, score and gradient statistics have the same asymptotic
distributional properties either under the null hypothesis or under a sequence of
local alternatives, i.e.~a sequence of Pitman
alternatives converging to the null hypothesis at a convergence rate $n^{-1/2}$.
On the other hand, up to an error of order $n^{-1}$ the corresponding criteria have the
same size properties but their local powers differ in the $n^{-1/2}$ term. A meaningful
comparison among the criteria can be performed by comparing
the nonnull asymptotic expansions to order $n^{-1/2}$ of the distribution functions
of these statistics under a sequence of Pitman alternatives. In this regard, we can benefit
from the work by \cite{Hayakawa1975}, \cite{HarrisPeers1980} and  \cite{LemFer2010a}.
\cite{Hayakawa1975} derived the nonnull asymptotic expansions up to order $n^{-1/2}$
for the densities of the likelihood ratio and Wald statistics, while an analogous result for the
score statistic was obtained by \cite{HarrisPeers1980}. Recently, the asymptotic expansion up
to order $n^{-1/2}$ for the density of the gradient statistic was derived
by \cite{LemFer2010a}. The expansions obtained by these authors are
extremely general but it can be very difficult or even impossible to
particularize their formulas for specific regression models.
As we shall see below, we have been able to apply their results for the $\mathcal{BS}$
regression model.

The rest of the paper is organized as follows. Section \ref{tests} briefly describes the
likelihood ratio, Wald, score and gradient tests.
In Section \ref{inference_BS} these tests are applied for testing hypotheses on
the parameters of the $\mathcal{BS}$ regression model. In Section \ref{main_result} we
obtain and compare the local powers of the tests. Monte Carlo simulation results on
the finite-sample performance of the tests are presented  and discussed in Section
\ref{MCsimulation}. Section \ref{application} contains an application to a real fatigue
data set. Finally, Section \ref{conclusions} discusses our main findings and closes
the paper with some conclusions.

\section{Background}\label{tests}


Consider a parametric model $f(\cdot; \bm{\theta})$ with corresponding
log-likelihood function $\ell(\bm{\theta})$, where $\bm{\theta} =
(\bm{\theta}_{1}^{\top}, \bm{\theta}_{2}^{\top})^{\top}$ is a $k$-vector
of unknown parameters. The dimensions of $\bm{\theta}_{1}$ and $\bm{\theta}_{2}$
are $k$ and $k-q$, respectively. Suppose the interest lies in testing the composite null hypothesis
$\mathcal{H}_{0}:\bm{\theta}_{2} = \bm{\theta}_{2}^{(0)}$
against $\mathcal{H}_{1}:\bm{\theta}_{2}\neq\bm{\theta}_{2}^{(0)}$,
where $\bm{\theta}_{2}^{(0)}$ is a specified vector.
Hence, $\bm{\theta}_{1}$ is a vector of nuisance parameters.
Let $\bm{U}_{\bm{\theta}}$ and $\bm{K}_{\bm{\theta}}$ denote the score
function and the Fisher information matrix for $\bm{\theta}$, respectively.
The partition for $\bm{\theta}$ induces the corresponding partitions
\[
\bm{U}_{\bm{\theta}} = (\bm{U}_{\bm{\theta}_1}^\top, \bm{U}_{\bm{\theta}_2}^\top)^\top,
\quad
\bm{K}_{\bm{\theta}} =
\begin{pmatrix}
\bm{K}_{\bm{\theta}11} & \bm{K}_{\bm{\theta}12} \\
\bm{K}_{\bm{\theta}21} & \bm{K}_{\bm{\theta}22}
\end{pmatrix},
\quad
\bm{K}_{\bm{\theta}}^{-1} =
\begin{pmatrix}
\bm{K}^{11} & \bm{K}^{12} \\
\bm{K}^{21} & \bm{K}^{22}
\end{pmatrix},
\]
where $\bm{K}_{\bm{\theta}}^{-1}$ is the inverse of $\bm{K}_{\bm{\theta}}$.

Let $\widehat{\bm{\theta}}=(\widehat{\bm{\theta}}_1^\top,\widehat{\bm{\theta}}_2^\top)^\top$ and
$\widetilde{\bm{\theta}}=(\widetilde{\bm{\theta}}_1^\top,\bm{\theta}_{2}^{(0)\top})^\top$
denote the maximum likelihood estimators of $\bm{\theta} =
(\bm{\theta}_{1}^{\top}, \bm{\theta}_{2}^{\top})^{\top}$ under
$\mathcal{H}_{1}$ and $\mathcal{H}_{0}$, respectively. The
likelihood ratio ($S_1$), Wald ($S_2$), score ($S_3$) and gradient ($S_4$) statistics for
testing $\mathcal{H}_{0}$ versus $\mathcal{H}_{1}$ are given by
\[
S_{1} = 2\bigl\{\ell(\widehat{\bm{\theta}}) - \ell(\widetilde{\bm{\theta}})\bigr\},
\qquad
S_{2} = (\widehat{\bm{\theta}}_{2} - \bm{\theta}_{2}^{(0)})^{\top}\widehat{\bm{K}}^{22^{-1}}
(\widehat{\bm{\theta}}_{2} - \bm{\theta}_{2}^{(0)}),
\]
\[
S_{3} = \widetilde{\bm{U}}_{\bm{\theta}_2}^\top\widetilde{\bm{K}}^{22}\widetilde{\bm{U}}_{\bm{\theta}_2},
\qquad
S_{4} = \widetilde{\bm{U}}_{\bm{\theta}_2}^\top(\widehat{\bm{\theta}}_{2} - \bm{\theta}_{2}^{(0)}),
\]
respectively, where $\widehat{\bm{K}}^{22}=\bm{K}^{22}(\widehat{\bm{\theta}})$,
$\widetilde{\bm{K}}^{22}=\bm{K}^{22}(\widetilde{\bm{\theta}})$ and
$\widetilde{\bm{U}}_{\bm{\theta}_2} = \bm{U}_{\bm{\theta}_2}(\widetilde{\bm{\theta}})$.
The limiting distribution of $S_1$, $S_2$, $S_3$ and $S_4$ is $\chi_{k-q}^2$ under
$\mathcal{H}_{0}$ and $\chi_{k-q,\lambda}^{2}$, i.e.~a noncentral chi-square
distribution with $k-q$ degrees of freedom and an appropriate noncentrality parameter $\lambda$,
under $\mathcal{H}_{1}$. The null hypothesis is rejected for a given nominal level, $\gamma$ say,
if the test statistic exceeds the upper $1-\gamma$ quantile of the $\chi_{k-q}^{2}$
distribution. Clearly, $S_{4}$ has a very simple form and does not involve
knowledge of the information matrix, neither expected nor observed,
and any matrix, unlike $S_{2}$ and $S_{3}$.

\section{Test statistics in the $\mathcal{BS}$ regression model}\label{inference_BS}

In what follows, we shall consider the tests which are based on the statistics
$S_1$, $S_2$, $S_3$ and $S_4$ in the class of $\mathcal{BS}$
regression models for testing a composite null hypothesis.
The log-likelihood function $\ell(\bm{\theta})$ for the vector parameter
$\bm{\theta}=(\bm{\beta}^{\top},\alpha)^{\top}$ from a random
sample $\bm{y}=(y_1,\ldots,y_n)^{\top}$ obtained from model~(\ref{eq1}), except
for constants, can be written as
\[
\ell(\bm{\theta})=\sum_{i=1}^{n}\log(\xi_{i1}) - \frac{1}{2}\sum_{i=1}^{n}\xi_{i2}^{2},
\]
where $\xi_{i1}=\xi_{i1}(\bm{\theta})= 2\alpha^{-1}\cosh([y_i-\mu_{i}]/2)$,
$\xi_{i2}=\xi_{i2}(\bm{\theta})=2\alpha^{-1}\sinh([y_i-\mu_{i}]/2)$ and
$\mu_{i} = \bm{x}_{i}^{\top}\bm{\beta}$,
for $i=1,\ldots,n$. It is assumed that the model matrix $\bm{X} = (\bm{x}_1,\ldots,\bm{x}_n)^{\top}$
has full column rank, i.e., rank$(\bm{X}) = p$. The score function and
the Fisher information matrix for $\bm{\theta}=(\bm{\beta}^{\top},\alpha)^{\top}$ are, respectively, given by
\[
\bm{U}_{\bm{\theta}} = 
(\bm{U}_{\bm{\beta}}^\top, U_{\alpha})^\top,\qquad
\bm{K}_{\bm{\theta}}=\diag\{\bm{K}_{\bm{\beta}},K_{\alpha}\},
\]
where $\bm{U}_{\bm{\beta}} = (1/2)\bm{X}^\top\bm{s}$,
$U_{\alpha} = -n/\alpha + (1/\alpha) \sum_{i=1}^{n}\xi_{i2}^{2}$,
$\bm{K}_{\bm{\beta}}=\psi(\alpha)(\bm{X}^{\top}\bm{X})/4$, $K_{\alpha} = 2n/\alpha^2$,
$\bm{s}=(s_1,\ldots,s_n)^\top$ with  $s_i =\xi_{i1}\xi_{i2}-\xi_{i2}/\xi_{i1}$ and
$\psi(\alpha) = 2 + 4/\alpha^2 - (\sqrt{2\pi}/\alpha)
\{1 - \mathtt{erf}(\sqrt{2}/\alpha)\}\exp(2/\alpha^2)$,
$\mathtt{erf}(\cdot)$ denoting the {\em error function}:
$\mathtt{erf}(x)=(2/\sqrt{\pi})\int_{0}^{x}\mathrm{e}^{-t^2}\mathrm{d}t$
\citep[see, for instance,][]{GR2007}.
From the block-diagonal form of $\bm{K}_{\bm{\theta}}$
we have that $\bm{\beta}$ and $\alpha$ are globally orthogonal
\citep{CoxReid87}.

The hypothesis of interest is
$\mathcal{H}_{0}: \bm{\beta}_{2} = \bm{\beta}_{2}^{(0)}$,
which will be tested against the alternative hypothesis
$\mathcal{H}_{1}:\bm{\beta}_{2}\neq\bm{\beta}_{2}^{(0)}$,
where $\bm{\beta}$ is partitioned as $\bm{\beta} = (\bm{\beta}_{1}^{\top},
\bm{\beta}_{2}^{\top})^{\top}$, with
$\bm{\beta}_{1} = (\beta_{1}, \dots,\beta_{q})^{\top}$ and
$\bm{\beta}_{2} = (\beta_{q+1},\dots,\beta_{p})^{\top}$. Here,
$\bm{\beta}_{2}^{(0)}$ is a fixed column vector of dimension $p-q$. The
partition for $\bm{\beta}$ induces the corresponding partitions
$\bm{U}_{\bm{\beta}} = (\bm{U}_{\bm{\beta}_1}^\top, \bm{U}_{\bm{\beta}_2}^\top)^\top$,
with $\bm{U}_{\bm{\beta}_1}=(1/2)\bm{X}_{1}^\top\bm{s}$ and
$\bm{U}_{\bm{\beta}_2}=(1/2)\bm{X}_{2}^\top\bm{s}$,
\[
\bm{K}_{\bm{\beta}} =
\begin{pmatrix}
\bm{K}_{\bm{\beta}11} & \bm{K}_{\bm{\beta}12} \\
\bm{K}_{\bm{\beta}21} & \bm{K}_{\bm{\beta}22}
\end{pmatrix} = \frac{\psi(\alpha)}{4}
\begin{pmatrix}
\bm{X}_{1}^\top\bm{X}_1 & \bm{X}_{1}^\top\bm{X}_2 \\
\bm{X}_{2}^\top\bm{X}_1 & \bm{X}_{2}^\top\bm{X}_2
\end{pmatrix},
\]
with the matrix $\bm{X}$ partitioned
as $\bm{X} = \begin{pmatrix}\bm{X}_{1} & \bm{X}_{2}\end{pmatrix}$.
The likelihood ratio, Wald, score and gradient statistics for testing $\mathcal{H}_{0}$
can be expressed, respectively, as
\[
S_{1} = 2\bigl\{\ell(\widehat{\bm{\theta}}) - \ell(\widetilde{\bm{\theta}})\bigr\},
\qquad
S_{2} = \frac{\psi(\widehat{\alpha})}{4}(\widehat{\bm{\beta}}_{2}
- \bm{\beta}_{2}^{(0)})^{\top}(\bm{R}^{\top}\bm{R})
(\widehat{\bm{\beta}}_{2} - \bm{\beta}_{2}^{(0)}),
\]
\[
S_{3} = \frac{1}{\psi(\widetilde{\alpha})}\widetilde{\bm{s}}^{\top}\bm{X}_{2}(\bm{R}^{\top}\bm{R})^{-1}
\bm{X}_{2}^{\top}\widetilde{\bm{s}},\qquad
S_{4} = \frac{1}{2}\widetilde{\bm{s}}^{\top}\bm{X}_{2}(\widehat{\bm{\beta}}_{2} - \bm{\beta}_{2}^{(0)}),
\]
where $\widehat{\bm{\theta}} = (\widehat{\bm{\beta}}_{1}^\top,
\widehat{\bm{\beta}}_{2}^\top,\widehat{\alpha})^\top$,
$\widetilde{\bm{\theta}} = (\widetilde{\bm{\beta}}_{1}^\top,
\bm{\beta}_{2}^{(0)\top},\widetilde{\alpha})^\top$,
$\bm{R} = \bm{X}_{2} - \bm{X}_{1}(\bm{X}_{1}^{\top}\bm{X}_{1})^{-1} \bm{X}_{1}^{\top}\bm{X}_{2}$
and $\widetilde{\bm{s}}=\bm{s}(\widetilde{\bm{\theta}})$.
The limiting distribution of all these statistics under $\mathcal{H}_{0}$ is $\chi_{p-q}^2$.
Notice that, unlike the Wald and score statistics, the gradient
statistic does note involve the error function.

Now, the problem under consideration is that of testing
a composite null hypothesis $\mathcal{H}_{0}:\alpha=\alpha^{(0)}$ against
$\mathcal{H}_{1}:\alpha\neq\alpha^{(0)}$, where $\alpha^{(0)}$ is a positive specified value
for $\alpha$, and $\bm{\beta}$ acts as a nuisance parameter.
The four statistics are expressed as follows:
\[
S_{1} = 2\bigl\{\ell(\widehat{\bm{\beta}},\widehat{\alpha}) -
\ell(\widetilde{\bm{\beta}}, \alpha^{(0)})\bigr\},
\qquad
S_{2} = 2n\biggl(\frac{\widehat{\alpha}-\alpha^{(0)}}{\widehat{\alpha}}\biggr)^2,
\]
\[
S_{3} = \frac{n(\bar{\xi}_{2}-1)^2}{2},
\qquad
S_{4} = n(\bar{\xi}_{2}-1)\biggl(\frac{\widehat{\alpha}-\alpha^{(0)}}{\alpha^{(0)}}\biggr),
\]
where $\bar{\xi}_{2} = \bar{\xi}_{2}(\widetilde{\bm{\theta}})
= (1/n)\sum_{i=1}^{n}\xi_{i2}^2(\widetilde{\bm{\theta}})$,
with $\widehat{\bm{\theta}} = (\widehat{\bm{\beta}}^\top, \widehat{\alpha})^\top$
and $\widetilde{\bm{\theta}} = (\widetilde{\bm{\beta}}^\top,\alpha^{(0)})^\top$
representing the unrestricted and restricted maximum
likelihood estimators of $\bm{\theta}$ under
$\mathcal{H}_{1}$ and $\mathcal{H}_{0}$, respectively.

\section{Local power}\label{main_result}

In this section we shall assume the following local alternative
hypothesis $\mathcal{H}_{1n}:\bm{\beta}_{2}=\bm{\beta}_{2}^{(0)} + \bm{\epsilon}$,
where $\bm{\epsilon} = (\epsilon_{q+1}, \ldots,\epsilon_{p})^{\top}$ with
$\epsilon_{r} = O(n^{-1/2})$ for $r=q+1,\ldots,p$. We follow the notation
in \cite{FerrariBotterCribari1997}. Let
\[
\bm{A} = \begin{pmatrix}
          \bm{A}_{\bm{\beta}} & \bm{0}\\
          \bm{0} & \kappa_{\alpha,\alpha}^{-1}
         \end{pmatrix},
\qquad
\bm{M} = \begin{pmatrix}
         \bm{M}_{\bm{\beta}} & \bm{0}\\
         \bm{0} & 0
         \end{pmatrix},
\]
where
\[
\bm{A}_{\bm{\beta}} = \begin{pmatrix}
                       \bm{K}_{\bm{\beta}11}^{-1} & \bm{0}\\
                       \bm{0} & \bm{0}
                      \end{pmatrix},
\qquad
\bm{M}_{\bm{\beta}} = \bm{K}_{\bm{\beta}}^{-1} - \bm{A}_{\bm{\beta}}.
\]
It then follows that $m_{r\alpha}= m_{\alpha r} = m_{\alpha\alpha} = 0$,
$a_{r\alpha} = a_{\alpha r} = 0$, for $r = 1, \ldots,p$, and
$a_{\alpha\alpha} = \kappa_{\alpha,\alpha}^{-1} =
\alpha^2/(2n)$, where $m_{r\alpha}$ and $a_{r\alpha}$ are the $(r,p+1)$
elements of the matrices $\bm{M}$ and $\bm{A}$, respectively, and $m_{\alpha\alpha}$ and
$a_{\alpha\alpha}$ are the $(p+1, p+1)$ elements of the matrices $\bm{M}$ and $\bm{A}$, respectively.
Additionally, let
\[
\bm{\epsilon}^{*} = \begin{pmatrix}
\bm{K}_{\bm{\beta}11}^{-1}\bm{K}_{\bm{\beta}12}\\
-\bm{I}_{p-q}\\
\bm{0}
\end{pmatrix}\bm{\epsilon},
\]
where $\bm{I}_{p-q}$ is a $(p-q)\times(p-q)$ identity matrix.

The nonnull distributions of the statistics $S_1$, $S_2$, $S_3$ and $S_4$
under Pitman alternatives for testing $\mathcal{H}_{0}:\bm{\beta}_{2}=\bm{\beta}_{2}^{(0)}$
in the $\mathcal{BS}$ regression model can be expressed as
\begin{equation*}\label{expansion}
\Pr(S_{i}\leq x) = G_{p-q,\lambda}(x) + \sum_{k=0}^{3}b_{ik}G_{p-q+2k,\lambda}(x) + O(n^{-1}),
\qquad i=1,2,3,4,
\end{equation*}
where $G_{m,\lambda}(x)$ is the cumulative distribution function of
a non-central chi-square variate with $m$ degrees of freedom
and non-centrality parameter $\lambda$. Here,
$\lambda = \bm{\epsilon}^{*\top}\bm{K}_{\bm{\theta}}\bm{\epsilon}^{*}$ and
the coefficients $b_{ik}$'s ($i=1,2,3,4$ and $k=0,1,2,3$) can be written as
\begin{align*}
b_{11} &= -\frac{1}{6}\sum_{r,s,t=1}^{p}(\kappa_{rst}
          - 2\kappa_{r,s,t})\epsilon_{r}^{*}\epsilon_{s}^{*}\epsilon_{t}^{*}
          -\frac{1}{2}\sum_{r,s,t=1}^{p}(\kappa_{rst} + 2\kappa_{r,st})a_{rs}\epsilon_{t}^{*}\\
          &-\frac{1}{2}\sum_{r=q+1}^{p}\sum_{s,t=1}^{p}(\kappa_{rst} + \kappa_{r,st})
          \epsilon_{r}\epsilon_{s}^{*}\epsilon_{t}^{*} -\frac{1}{2}\sum_{t=1}^{p}(\kappa_{\alpha\alpha t}+
          2\kappa_{\alpha,\alpha t})\kappa_{\alpha,\alpha}^{-1}\epsilon_{t}^*,
\end{align*}
\[
b_{12} = -\frac{1}{6}\sum_{r,s,t=1}^{p}\kappa_{r,s,t}\epsilon_{r}^{*}\epsilon_{s}^{*}\epsilon_{t}^{*},\quad
b_{13} = 0,
\]
\begin{align*}
b_{21} &= -\frac{1}{2}\sum_{r,s,t=1}^{p}(\kappa_{rst} + 2\kappa_{r,st})\epsilon_{r}^{*}\epsilon_{s}^{*}\epsilon_{t}^{*}
          +\sum_{r,s,t=1}^{p}\kappa_{r,st}m_{rs}\epsilon_{t}^{*}
         -\frac{1}{2}\sum_{r,s,t=1}^{p}(\kappa_{rst} + 2\kappa_{r,st})\kappa^{r,s}\epsilon_{t}^{*}\\
       &  -\frac{1}{2}\sum_{r=q+1}^{p}\sum_{s,t=1}^{p}(\kappa_{rst} + \kappa_{r,st})
        \epsilon_{r}\epsilon_{s}^{*}\epsilon_{t}^{*}-\frac{1}{2}\sum_{t=1}^{p}(\kappa_{\alpha\alpha t}+
          2\kappa_{\alpha,\alpha t})\kappa_{\alpha,\alpha}^{-1}\epsilon_{t}^*,
\end{align*}
\[
b_{22} = \frac{1}{2}\sum_{r,s,t=1}^{p}\kappa_{r,st}\epsilon_{r}^{*}\epsilon_{s}^{*}\epsilon_{t}^{*}
        + \frac{1}{2}\sum_{r,s,t=1}^{p}\kappa_{rst}m_{rs}\epsilon_{t}^{*},\quad
b_{23} = \frac{1}{6}\sum_{r,s,t=1}^{p}\kappa_{rst}\epsilon_{r}^{*}\epsilon_{s}^{*}\epsilon_{t}^{*},
\]
\begin{align*}
b_{31} &= -\frac{1}{6}\sum_{r,s,t=1}^{p}(\kappa_{rst} - 2\kappa_{r,s,t})
          \epsilon_{r}^{*}\epsilon_{s}^{*}\epsilon_{t}^{*}
          +\frac{1}{2}\sum_{r,s,t=1}^{p}\kappa_{r,s,t}m_{rs}\epsilon_{t}^{*}
         -\frac{1}{2}\sum_{r,s,t=1}^{p}(\kappa_{rst} + 2\kappa_{r,st})a_{rs}\epsilon_{t}^{*}\\
        &  -\frac{1}{2}\sum_{r=q+1}^{p}\sum_{s,t=1}^{p}(\kappa_{rst} + \kappa_{r,st})
          \epsilon_{r}\epsilon_{s}^{*}\epsilon_{t}^{*}-\frac{1}{2}\sum_{t=1}^{p}(\kappa_{\alpha\alpha t}+
          2\kappa_{\alpha,\alpha t})\kappa_{\alpha,\alpha}^{-1}\epsilon_{t}^*,
\end{align*}
\[
b_{32} = -\frac{1}{2}\sum_{r,s,t=1}^{p}\kappa_{r,s,t}m_{rs}\epsilon_{t}^{*},\quad
b_{33} = -\frac{1}{6}\sum_{r,s,t=1}^{p}\kappa_{r,s,t}\epsilon_{r}^{*}\epsilon_{s}^{*}\epsilon_{t}^{*},
\]
\begin{align*}
b_{41} &= \frac{1}{4}\sum_{r,s,t=1}^{p}\kappa_{rst}\kappa^{r,s}\epsilon_{t}^{*}
        -\frac{1}{2}\sum_{r,s,t=1}^{p}(\kappa_{rst} + 2\kappa_{r,st})
         \epsilon_{r}^{*}\epsilon_{s}^{*}\epsilon_{t}^{*}
        -\frac{1}{4}\sum_{r,s,t=1}^{p}(4\kappa_{r,st} + 3\kappa_{r,st})a_{rs}\epsilon_{t}^{*}\\
       & -\frac{1}{2}\sum_{r=q+1}^{p}\sum_{s,t=1}^{p}(\kappa_{rst} + \kappa_{r,st})
        \epsilon_{r}\epsilon_{s}^{*}\epsilon_{t}^{*}-\frac{1}{2}\sum_{t=1}^{p}(\kappa_{\alpha\alpha t}+
          2\kappa_{\alpha,\alpha t})\kappa_{\alpha,\alpha}^{-1}\epsilon_{t}^*,
\end{align*}
\[
b_{42} = -\frac{1}{4}\sum_{r,s,t=1}^{p}\kappa_{rst}m_{rs}\epsilon_{t}^{*}
        + \frac{1}{4}\sum_{r,s,t=1}^{p}(\kappa_{rst} + 2\kappa_{r,st})
          \epsilon_{r}^{*}\epsilon_{s}^{*}\epsilon_{t}^{*}
\qquad
b_{43} = -\frac{1}{12}\sum_{r,s,t=1}^{p}\kappa_{r,s,t}\epsilon_{r}^{*}\epsilon_{s}^{*}\epsilon_{t}^{*},
\]
where the $\kappa$'s are defined as
$\kappa_{rs} = \Es(\partial^2\ell(\bm{\theta})/\partial\beta_r\partial\beta_s)$,
$\kappa_{rst} = \Es(\partial^3\ell(\bm{\theta})/\partial\beta_r\partial\beta_s\partial\beta_t)$,
$\kappa_{r,st} = \Es\{(\partial\ell(\bm{\theta})/\partial\beta_r)
(\partial^2\ell(\bm{\theta})/\partial\beta_s\partial\beta_t)\}$,
$\kappa_{\alpha\alpha t} = \Es(\partial^3\ell(\bm{\theta})/\partial\alpha^2\partial\beta_t)$, etc.,
and $\kappa^{r,s}$ is the ($r,s$)th element of inverse of $\bm{K}_{\bm{\beta}}$.
The coefficients $b_{i0}$ are obtained from $b_{i0} = -(b_{i1} + b_{i2} + b_{i3})$, for $i=1,2,3,4$.

After some algebra, it is possible to show that, in the $\mathcal{BS}$  regression model,
\[
\kappa_{rst} = \kappa_{r,st} = \kappa_{r,s,t} =
\kappa_{\alpha\alpha t}= \kappa_{\alpha\alpha, t} = 0, \qquad r,s,t = 1,\ldots,p.
\]
Therefore, $b_{ik} = 0$ for $i=1,2,3,4$ and $k=0,1,2,3$, and we can write
\[
\Pr(S_{i}\leq x) = G_{p-q,\lambda}(x) + O(n^{-1}),\qquad i=1,2,3,4.
\]
This is a very interesting result, which implies that the likelihood ratio, score, Wald
and gradient tests for testing the composite null hypothesis 
$\mathcal{H}_{0}:\bm{\beta}_{2}=\bm{\beta}_{2}^{(0)}$
have exactly the same local power up to an error of order $n^{-1}$.

We now turn to the problem of testing hypotheses on  $\alpha$, the shape parameter.
The nonnull asymptotic distributions of the statistics $S_1$, $S_2$, $S_3$
and $S_4$ for testing $\mathcal{H}_{0}:\alpha=\alpha^{(0)}$
under the local alternative $\mathcal{H}_{1n}:\alpha=\alpha^{(0)}+\epsilon$,
where $\epsilon=\alpha-\alpha^{(0)}$ is assumed to be $O(n^{-1/2})$, is
\[
\Pr(S_{i}\leq x) = G_{1,\lambda}(x) + \sum_{k=0}^{3}b_{ik}G_{1+2k,\lambda}(x) + O(n^{-1}),
\qquad i=1,2,3,4,
\]
with $\lambda = 2n\epsilon^2/\alpha^2$.
The $b_{ik}$'s for the test of $\mathcal{H}_{0}:\alpha=\alpha^{(0)}$ are easy to obtain
and are given by
$b_{11} = (\kappa_{\alpha\alpha\alpha} - 2\kappa_{\alpha,\alpha,\alpha})\epsilon^3/6
+\sum_{r,s=1}^{p}(\kappa_{rs\alpha} + 2\kappa_{r,s\alpha})\kappa^{r,s}\epsilon/2
-(\kappa_{\alpha\alpha\alpha} + \kappa_{\alpha,\alpha\alpha})\epsilon^3/2$,
$b_{12} = \kappa_{\alpha,\alpha,\alpha}\epsilon^3/6$, $b_{13} = 0$,
$b_{21} = (\kappa_{\alpha\alpha\alpha} + 2\kappa_{\alpha,\alpha\alpha})\epsilon^3/2
-\kappa_{\alpha,\alpha\alpha}\kappa^{\alpha,\alpha}\epsilon
+\sum_{r,s=1}^{p}(\kappa_{rs\alpha} + 2\kappa_{r,s\alpha})\kappa^{r,s}\epsilon/2
+(\kappa_{\alpha\alpha\alpha} + 2\kappa_{\alpha,\alpha\alpha})\kappa^{\alpha,\alpha}\epsilon/2
-(\kappa_{\alpha\alpha\alpha} + \kappa_{\alpha,\alpha\alpha})\epsilon^3/2$,
$b_{22} = -(\kappa_{\alpha,\alpha\alpha}\epsilon^3
+ \kappa_{\alpha\alpha\alpha}\kappa^{\alpha,\alpha}\epsilon)/2$,
$b_{23} = -\kappa_{\alpha\alpha\alpha}\epsilon^3/6$,
$b_{31} = (\kappa_{\alpha\alpha\alpha} - 2\kappa_{\alpha,\alpha,\alpha})\epsilon^3/6
-\kappa_{\alpha,\alpha,\alpha}\kappa^{\alpha,\alpha}\epsilon/2
+\sum_{r,s=1}^{p}(\kappa_{rs\alpha} + 2\kappa_{r,s\alpha})\kappa^{r,s}\epsilon/2
-(\kappa_{\alpha\alpha\alpha} + \kappa_{\alpha,\alpha\alpha})\epsilon^3/2$,
$b_{32} = \kappa_{\alpha,\alpha,\alpha}\kappa^{\alpha,\alpha}\epsilon/2$,
$b_{33} = \kappa_{\alpha,\alpha,\alpha}\epsilon^3/6$,
$b_{41} = -\sum_{r,s=1}^{p}\kappa_{rs\alpha}\kappa^{r,s}\epsilon/4
-\kappa_{\alpha\alpha\alpha}\kappa^{\alpha,\alpha}\epsilon/4
+\kappa_{\alpha,\alpha\alpha}\epsilon^3/2
+\sum_{r,s=1}^{p}(4\kappa_{r,s\alpha} + 3\kappa_{rs\alpha})\kappa^{r,s}\epsilon/4$,
$b_{42} = \kappa_{\alpha\alpha\alpha}\kappa^{\alpha, \alpha}\epsilon/4
-(\kappa_{\alpha\alpha\alpha} + 2\kappa_{\alpha,\alpha\alpha})\epsilon^3/4$ and
$b_{43} = \kappa_{\alpha\alpha\alpha}\epsilon^3/12$,
where $\kappa^{\alpha,\alpha} = \alpha^2/(2n)$.
The coefficients $b_{i0}$ are obtained from $b_{i0} = -(b_{i1} + b_{i2} + b_{i3})$, for $i=1,2,3,4$.
We have that
\[
\kappa_{\alpha\alpha\alpha} = \frac{10n}{\alpha^3},\quad
\kappa_{\alpha,\alpha\alpha} = -\frac{6n}{\alpha^3},\quad
\kappa_{\alpha,\alpha,\alpha} = \frac{8n}{\alpha^3}, \quad
\kappa_{rs\alpha} = -\kappa_{r,s\alpha} = \frac{(2+\alpha^2)}
{\alpha^3}\sum_{i=1}^{n}x_{ir}x_{is},
\]
for $r,s=1,\ldots,p$. After some algebra the coefficients $b_{ik}$ reduce to
\[
b_{11}=-\frac{3n\epsilon^3}{\alpha^3}-\frac{2p(2+\alpha^2)\epsilon}{\alpha^3\psi(\alpha)},\quad
b_{12} = \frac{4n\epsilon^3}{3\alpha^3},\quad
b_{13} = 0,\quad
b_{21}=-\frac{3n\epsilon^3}{\alpha^3}+\frac{5\epsilon}{2\alpha}-
\frac{2p(2+\alpha^2)\epsilon}{\alpha^3\psi(\alpha)},\quad
\]
\[
b_{22} = \frac{3n\epsilon^3}{\alpha^3} -\frac{5\epsilon}{2\alpha},\quad
b_{23} = -\frac{5n\epsilon^3}{3\alpha^3},\quad
b_{31}=-\frac{3n\epsilon^3}{\alpha^3}-\frac{2\epsilon}{\alpha}-
\frac{2p(2+\alpha^2)\epsilon}{\alpha^3\psi(\alpha)},\quad
b_{32} = \frac{2\epsilon}{\alpha},
\]
\[
b_{33} = \frac{4n\epsilon^3}{3\alpha^3}, \qquad
b_{41}=-\frac{3n\epsilon^3}{\alpha^3}-\frac{5\epsilon}{4\alpha}-
\frac{2p(2+\alpha^2)\epsilon}{\alpha^3\psi(\alpha)},\quad
b_{42} = \frac{5\epsilon}{4\alpha} + \frac{n\epsilon^3}{2\alpha^3}, \quad
b_{43} = \frac{5n\epsilon^3}{6\alpha^3}.
\]
It should be noticed that the above expressions depend on the model only through
$\alpha$ and the rank of the model matrix $\bm{X}$; they do not
involve the unknown parameter $\bm{\beta}$.

We now present an analytical comparison among the local powers of the four tests for testing
the null hypothesis $\mathcal{H}_{0}:\alpha=\alpha^{(0)}$.
Let $\Pi_{i}$ be the power function, up to order $n^{-1/2}$, of the test that uses
the statistic $S_{i}$, for $i=1,2,3,4$.
We have
\begin{equation}\label{diff_power}
\Pi_{i} - \Pi_{j} = \sum_{k=0}^{3}(b_{jk} - b_{ik})G_{1+2k,\lambda}(x),
\end{equation}
for $i\neq j$. It is well known that
\begin{equation}\label{diff_G}
G_{m,\lambda}(x) - G_{m+2,\lambda}(x) = 2g_{m+2,\lambda}(x),
\end{equation}
where $g_{\nu,\lambda}(x)$ is the probability density
function of a non-central chi-square random variable
with $\nu$ degrees of freedom and non-centrality parameter $\lambda$.
From~(\ref{diff_power}) and (\ref{diff_G}), we have
\[
\Pi_{1}-\Pi_{2} = \frac{5\epsilon}{\alpha} g_{5,\lambda}(x)
+ \frac{10n\epsilon^3}{3\alpha^3} g_{7,\lambda}(x),
\quad
\Pi_{1}-\Pi_{3} = -\biggl\{\frac{4\epsilon}{\alpha} g_{5,\lambda}(x)
+ \frac{8n\epsilon^3}{3\alpha^3} g_{7,\lambda}(x)\biggr\},
\]
\[
\Pi_{1}-\Pi_{4} = -\biggl\{\frac{5\epsilon}{2\alpha} g_{5,\lambda}(x)
+ \frac{5n\epsilon^3}{3\alpha^3} g_{7,\lambda}(x)\biggr\},
\quad
\Pi_{3}-\Pi_{4} = \frac{3\epsilon}{2\alpha} g_{5,\lambda}(x)
+ \frac{n\epsilon^3}{\alpha^3} g_{7,\lambda}(x).
\]
Hence, we arrive at the following inequalities:
$\Pi_{3} > \Pi_{4} > \Pi_{1} > \Pi_{2}$ if $\alpha > \alpha^{(0)}$,  and
$\Pi_{3} < \Pi_{4} < \Pi_{1} < \Pi_{2}$ if $\alpha < \alpha^{(0)}$.

The most important finding obtained so far is that
the likelihood ratio, score, Wald and gradient tests
for testing the null hypothesis $\mathcal{H}_{0}:\bm{\beta}_{2}=\bm{\beta}_{2}^{(0)}$
share the same null size and local power up to an error of order $n^{-1}$.
To this order of approximation the null distribution of the four statistics
is $\chi^2_{p-q}$. Therefore, if the sample size is large, type I error
probabilities of all the tests
do not significantly deviate from the true nominal level, and
their powers are approximately equal for alternatives that are close to the
null hypothesis.

The natural question now is how these tests perform when the sample size is
small or of moderate size, and which one is the most reliable.
In the next section, we shall use Monte Carlo simulations
to put some light on this issue.

\section{Finite-sample performance}\label{MCsimulation}

In this section we shall present the results of a Monte Carlo simulation
in which we evaluate the finite sample performance of the likelihood ratio,
Wald, score and gradient tests. The simulations were based on the model
\[
y_{i} = \beta_{1}x_{i1} + \beta_{2}x_{i2} + \cdots + \beta_{p}x_{ip} + \varepsilon_{i},
\]
where $x_{i1} = 1$ and $\varepsilon_{i}\sim\mathcal{SN}(\alpha, 0, 2)$, $i = 1,\ldots, n$.
The covariate values were selected as random draws from the uniform
$\mathcal{U}(0,1)$ distribution and for fixed $n$ those values were kept constant
throughout the experiment. The number of Monte Carlo replications was
15,000, the nominal levels of the tests were $\gamma$ = 10\%, 5\% and 1\%,
and all simulations were performed using the {\sf Ox} matrix
programming language \citep{DcK2007}. {\sf Ox} is freely distributed for academic
purposes and available at {\tt http://www.doornik.com}.
All log-likelihood maximizations with respect to $\bm{\beta}$ and $\alpha$ were carried out using
the BFGS quasi-Newton method with analytic first derivatives
through {\tt MaxBFGS} subroutine.
This method is generally regarded as the best-performing
nonlinear optimization method \citep[][p.~199]{Mittelhammer-et-al-2000}.
The initial values in the iterative BFGS scheme were
$\widetilde{\bm{\beta}} = (\bm{X}^{\top}\bm{X})^{-1}\bm{X}^{\top}\bm{y}$
for $\bm{\beta}$ and $\sqrt{\widetilde{\alpha}^2}$
for $\alpha$, where
\[
\widetilde{\alpha}^2 = \frac{4}{n}\sum_{i=1}^{n}\sinh^{2}\biggr(\frac{y_{i} -
                        \bm{x}_{i}^{\top}\widetilde{\bm{\beta}}}{2}\biggl).
\]

First, the null hypothesis is $\mathcal{H}_{0}:\beta_{p-1} =\beta_{p} = 0$,
which is tested against a two-sided alternative. The sample size is $n=25$,
$\alpha = 0.5,  1.0$ and $p = 3, 4, \ldots, 7$. The values of the
response were generated using $\beta_{1} = \cdots=\beta_{p-2} = 1$.
The null rejection rates of the four tests are presented in Table~\ref{tab1}.
It is evident that the likelihood ratio ($S_1$) and Wald ($S_2$) tests are
markedly liberal, more so as the number of regressors increases.
The score ($S_3$) and gradient ($S_4$) tests  are also liberal in most of
the cases, but much less size distorted than the likelihood ratio
and Wald tests in all cases. For instance, when $\alpha=0.5$, $p=6$ and $\gamma = 5\%$, the rejection rates
are 10.12\% ($S_1$), 12.77\% ($S_2$),  7.12\% ($S_3$) and 7.32\% ($S_4$).
It is noticeable that the score test is much less liberal than the
likelihood ratio and Wald tests and slightly less liberal than the
gradient test. The score and gradient tests are slightly conservative in some cases.
Additionally, the Wald test is much more liberal than the other tests.
Similar results hold for $\alpha=1.0$.
Table~\ref{tab2} reports results for $\alpha = 0.5$ and $p=5$ and sample sizes ranging
from 15 to 200.
As expected, the null rejection rates of all the tests approach the
corresponding nominal levels as the sample size grows. Again,
the score and gradient tests present the best performances.
\begin{table}[!htp]
\begin{center}
\caption{Null rejection rates (\%); $\alpha$ = 0.5 and 1.0, with $n = 25$.}\label{tab1}
\begin{tabular}{c c c c c| c c c c| c c c c  }\hline
  & \multicolumn{12}{c}{$\alpha = 0.5$} \\\cline{2-13}
  & \multicolumn{4}{c|}{$\gamma = 10\%$} & \multicolumn{4}{c|}{$\gamma = 5\%$}
  & \multicolumn{4}{c}{$\gamma = 1\%$}\\\cline{2-13}
  $p$   & $S_1$  & $S_2$  & $S_3$ & $S_4$
        & $S_1$  & $S_2$  & $S_3$ & $S_4$
        & $S_1$  & $S_2$  & $S_3$ & $S_4$\\\hline
    3   & 13.10& 15.73& 10.20& 10.46&  7.25&  9.41&  4.78&  4.97& 1.71& 3.32& 0.48& 0.57 \\
    4   & 14.37& 17.05& 11.51& 11.66&  8.04& 10.61&  5.35&  5.51& 2.01& 3.70& 0.73& 0.77 \\
    5   & 15.69& 18.64& 12.68& 12.99&  8.87& 11.86&  5.97&  6.21& 2.56& 4.37& 1.02& 1.09 \\
    6   & 17.13& 20.04& 13.79& 14.13& 10.12& 12.77&  7.12&  7.32& 2.83& 4.99& 1.01& 1.05 \\
    7   & 19.04& 22.07& 15.36& 15.73& 11.30& 14.57&  7.86&  8.15& 3.51& 5.90& 1.47& 1.57 \\ \hline
  & \multicolumn{12}{c}{$\alpha = 1.0$} \\\cline{2-13}
  & \multicolumn{4}{c|}{$\gamma = 10\%$} & \multicolumn{4}{c|}{$\gamma = 5\%$}
  & \multicolumn{4}{c}{$\gamma = 1\%$}\\\cline{2-13}
  $p$   & $S_1$  & $S_2$  & $S_3$ & $S_4$
        & $S_1$  & $S_2$  & $S_3$ & $S_4$
        & $S_1$  & $S_2$  & $S_3$ & $S_4$\\\hline
    3   & 12.69& 16.27& 10.18&  9.20&  6.82&  9.91&  4.00&  4.71& 1.66& 3.46& 0.41& 0.55 \\
    4   & 13.94& 17.60&  9.84& 11.07&  7.64& 10.87&  4.23&  5.07& 1.89& 3.81& 0.55& 0.75 \\
    5   & 15.91& 20.41& 11.15& 12.69&  8.76& 12.53&  4.92&  6.05& 2.35& 4.59& 0.64& 0.93 \\
    6   & 16.74& 20.65& 12.28& 13.86&  9.77& 13.61&  5.77&  6.89& 2.71& 5.39& 0.84& 1.09 \\
    7   & 18.67& 22.90& 13.49& 15.51& 11.27& 15.59&  6.77&  8.03& 3.55& 6.59& 1.00& 1.47 \\ \hline
\end{tabular}
\end{center}
\end{table}
\begin{table}[!htp]
\begin{center}
\caption{Null rejection rates (\%); $\alpha = 0.5$,
        $p = 5$ and different sample sizes.}\label{tab2}
\begin{tabular}{c c c c c| c c c c| c c c c  }\hline
  & \multicolumn{4}{c|}{$\gamma = 10\%$} & \multicolumn{4}{c|}{$\gamma = 5\%$}
  & \multicolumn{4}{c}{$\gamma = 1\%$}\\\cline{2-13}
  $n$   & $S_1$  & $S_2$  & $S_3$ & $S_4$
        & $S_1$  & $S_2$  & $S_3$ & $S_4$
        & $S_1$  & $S_2$  & $S_3$ & $S_4$\\\hline
   15   & 21.37& 26.38& 15.41& 15.85& 13.41& 19.02& 7.47& 7.89& 4.81&  9.47& 0.88& 0.97 \\
   20   & 17.65& 21.24& 13.73& 14.11& 10.81& 14.41& 6.83& 7.10& 3.41&  6.20& 0.99& 1.07 \\
   30   & 14.71& 16.95& 12.39& 12.50&  8.24& 10.45& 5.84& 6.07& 2.17&  3.60& 0.88& 0.93 \\
   40   & 13.49& 15.38& 11.73& 11.91&  7.25&  8.93& 5.47& 5.66& 1.65&  2.57& 0.88& 0.90 \\
   50   & 12.93& 14.27& 11.54& 11.80&  6.87&  8.21& 5.69& 5.73& 1.54&  2.14& 0.96& 1.04 \\
  100   & 11.77& 12.30& 10.95& 11.01&  5.79&  6.28& 5.25& 5.31& 1.23&  1.51& 0.93& 0.97 \\
  200   & 10.89& 11.11& 10.49& 10.57&  5.61&  5.90& 5.33& 5.38& 1.10&  1.25& 1.00& 0.99 \\ \hline
\end{tabular}
\end{center}
\end{table}

\begin{figure}
\centering
\includegraphics[width=10cm, height=7.5cm]{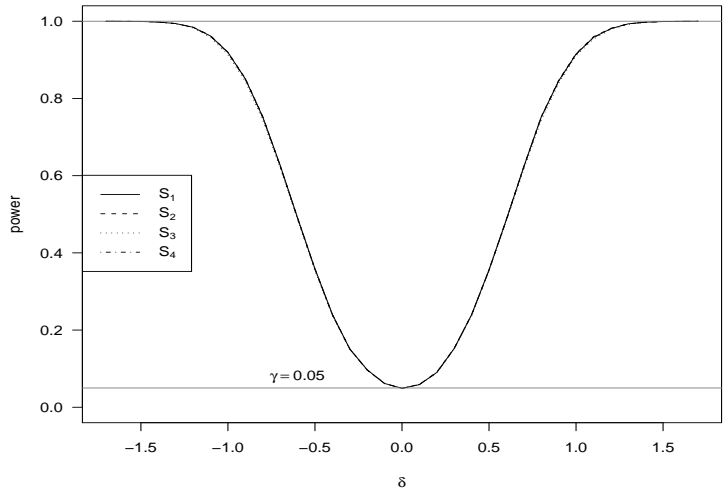}
\caption{Power of four tests: $n = 25$, $p = 4$, $\alpha = 0.5$ and $\gamma=5\%$.}\label{figpower}
\end{figure}

We now turn to the finite-sample power properties of the four tests.
The simulation results above show that the tests have different sizes when one
uses their asymptotic $\chi^2$ distribution in small and moderate-sized
samples. In evaluating the power of these tests,
it is important to ensure that they all have the correct size under the null
hypothesis. To overcome this difficulty, we used 500,000 Monte Carlo simulated
samples, drawn under the null hypothesis, to estimate the exact critical value of
each test for the chosen nominal level. We set $n = 25$, $p = 4$, $\alpha = 0.5$ and $\gamma=5\%$.
For the power simulations we computed the rejection rates under the alternative
hypothesis $\beta_3=\beta_4=\delta$, for $\delta$ ranging from $-2.0$ to $2.0$.
Figure~\ref{figpower} shows that the power curves of the four tests are indistinguishable
from each other. As expected, the powers of the tests approach 1 as $|\delta|$
grows. Power simulations carried out for other values of $n$, $p$ and
$\gamma$ showed a similar pattern.

Overall, in small to moderate-sized samples the best performing tests are the
score and the gradient tests. They are less size distorted than the other two
and are as powerful as the others.

We also performed Monte Carlo simulations considering hypothesis testing on $\alpha$.
To save space, the results are not shown. The score and gradient tests exhibited superior
behavior than the likelihood ratio and Wald tests. For example, when $n=35$, $p=4$, $\gamma = 5\%$ and $\mathcal{H}_{0}:\alpha = 0.5$, we obtained the following null rejection rates: 9.99\%
($S_1$), 15.89\% ($S_2$), 5.29\% ($S_3$) and 6.99\% ($S_4$).
Again, the best performing tests are the score and gradient tests.

\section{Application}\label{application}

This application focuses on modeling the die lifetime ($T$) in
the process of metal extrusion. The data were taken from  \cite{Lepadatu-et-al-2005}.
According to the authors, ``the estimation of tool life (fatigue
life) in the extrusion operation is important for scheduling tool changing
times, for adaptive process control and for tool cost evaluation.''
The authors noted that ``die fatigue cracks are caused by the repeat application
of loads which individually would be too small to cause failure.''
The $\mathcal{BS}$ regression model is then appealing in this context since
the main motivation for the $\mathcal{BS}$ distribution is
the fatigue failure time due to propagation of an initial crack.

We consider the following regression model:
\[
y_{i} = \beta_{1}x_{1i} + \beta_{2}x_{2i} + \beta_{3}x_{3i} + \beta_{4}x_{4i} + \beta_{5}x_{2i}x_{3i}
        + \beta_{6}x_{2i}x_{4i} + \beta_{7}x_{3i}x_{4i} + \varepsilon_{i},
\]
where $y_{i} = \log(T_{i})$, $\varepsilon_{i}\sim\mathcal{SN}(\alpha, 0, 2)$,
$x_{1i}=1$ and the covariates are $x_{2i}$ (friction coefficient),
$x_{3i}$ (angle of the die) and $x_{4i}$ (work temperature), for $i=1,\ldots,15$.
We wish to test the significance of the interaction effects, i.e., the interest lies in
testing $\mathcal{H}_{0}: \beta_{5} = \beta_{6} = \beta_{7} = 0$.
The likelihood ratio, Wald, score and gradient test statistics
equal $6.387$ ($p$-value: 0.094), $8.039$ ($p$-value: 0.045), $5.144$ ($p$-value: 0.162)
and $5.206$ ($p$-value: 0.157), respectively.
Hence, the null hypothesis is rejected at the 10\% nominal level when inference is based on the
likelihood ratio or the Wald test, but the opposite decision is reached when either the
score or the gradient test is used. Recall that our simulation results indicated that the
likelihood ratio and Wald tests are markedly liberal in small samples (here, $n=15$),
which leads us to mistrust the inference delivered by the likelihood ratio and Wald
tests. Therefore, we removed the interaction effects from the model as indicated by
the score and gradient tests.

The model containing only main effects is
$y_{i} = \beta_{1} + \beta_{2}x_{2i} + \beta_{3}x_{3i} + \beta_{4}x_{4i} + \varepsilon_{i},$
for $i=1,\dots,15$.
The null hypothesis  $\mathcal{H}_{0}: \beta_{3} = 0$ is strongly rejected
by the four tests at the usual significance levels.
All tests also suggest the individual and joint exclusions
of the friction coefficient and angle of the die
from the model. Hence, we end up with the regression model
$y_{i} = \beta_{1} + \beta_{4}x_{4i} + \varepsilon_{i}$, for $i=1,\ldots,15$.
The maximum likelihood estimates of the parameters are (standard errors in parentheses):
$\widehat{\beta}_{1} = 6.2453\, (0.326)$, $\widehat{\beta}_{4} = 0.0052\, (0.001)$
and $\widehat{\alpha} = 0.2039\, (0.037)$.

\section{Discussion}\label{conclusions}

The $\mathcal{BS}$ regression model is becoming increasingly popular
in lifetime analyses and reliability studies.
In this paper, we dealt with the issue of performing hypothesis testing
concerning the parameters of this model. We considered the three
classic tests, likelihood ratio, Wald and score tests, and a recently
proposed test, the gradient test. For the discussion that follows,
let us concentrate on tests regarding the regression parameters,
which are, in general, of primary interest.

The four tests have the same
distribution, under either the null hypothesis
or a sequence of local alternatives, up to an error of order $n^{-1}$, as we showed.
Our Monte Carlo simulation study added some important information. It
revealed that the likelihood ratio and the Wald tests can be remarkably
oversized if the sample is small. The score and
the gradient tests are clearly much less size distorted than the other
two tests. Our power simulations suggested that all the four tests have
similar power properties when estimated correct critical values
are used. Overall, this is an indication that the score and the gradient
tests should be prefered.

At this point, a discussion on small-sample corrections for the classic tests is in
order. A Bartlett correction for the likelihood ratio statistic and a Bartlett-type
correction for the score statistic were derived in \cite{Lemonte-et-al-2010}
and \cite{LemFer2010b}, respectively; see also Cordeiro and Ferrari~(1991).
The corrected statistics have the following interesting properties:
(i) the uncorrected and corrected statistics have the same asymptotic
distribution under the null hypothesis; (ii) the order of the error of the approximation
for the distribution of the test statistics by $\chi^2$ is smaller for
the corrected statistics than for the uncorrected statistics;
(iii) the corrections have no effect on the $n^{-1/2}$
term of the local power of the corresponding tests;
(iv) simulation results in \cite{Lemonte-et-al-2010} and \cite{LemFer2010b}
show that these corrections reduce the size distortion of the tests
and that the best performing test in small and moderate-sized
samples is the test which uses the corrected score statistic.
Therefore, the Bartlett-type-corrected score test is
an excellent alternative to the tests under consideration in the present paper.
The slight disadvantage of such an alternative is the extra computational
burden involved in computing the Bartlett-type correction.

We computed the corrected versions of likelihood ratio and score statistics for the
hypotheses tested in the real data application presented in Section \ref{application}.
Recall that the hypothesis of no interaction effects is rejected by the likelihood
ratio and Wald tests, but not rejected by the score and gradient tests. It is noteworthy
that the decision reached by either the corrected likelihood ratio test or the corrected score
test is in agreement with that obtained by the later two tests and in disagreement with
the likelihood ratio and Wald tests, which tend to reject the null hypothesis much more often than
indicated by the significance level.

Finally, our overall recommendations for practitioners when performing testing inference
in $\mathcal{BS}$ regressions are as follows. The score test or the gradient test
should be prefered as both perform better than the likelihood ratio and Wald tests
in small and moderate-sized samples. While the gradient test is a little more liberal
than the score test, it is easier to calculate. The Bartlett-type corrected
score test is a further better option although it requires a small extra computational effort.

\section*{Acknowledgments}

We gratefully acknowledge grants from FAPESP and CNPq (Brazil).


{\small
\begin{thebibliography}{99}

\bibitem[Balakrishnan et al.(2007)]{Balakrishnan-et-al-2007}
Balakrishnan, N., Leiva, V., L\'opez, J. (2007).
Acceptance sampling plans from truncated life tests from generalized Birnbaum--Saunders
distribution. {\em Communications in Statistics -- Simulation and Computation} {\bf 36}, 643--656.

\bibitem[Birnbaum and Saunders(1969a)]{BSa1969a}
Birnbaum, Z.W., Saunders, S.C. (1969a).
\newblock A new family of life distributions.
\newblock {\em Journal of Applied Probability\/} {\bf 6}, 319--327.

\bibitem[Birnbaum and Saunders(1969b)]{BSa1969b}
Birnbaum, Z.W., Saunders, S.C. (1969b).
\newblock Estimation for a family of life distributions with applications to fatigue.
\newblock {\em Journal of Applied Probability\/} {\bf 6}, 328--377.

\bibitem[Cordeiro and Ferrari(1991)]{CordeiroFerrari1991}
Cordeiro, G.M., Ferrari, S.L.P. (1991).
\newblock A modified score test statistic having chi-squared distribution to order $n^{-1}$.
\newblock {\em Biometrika\/} {\bf 78}, 573--582.


\bibitem[Cox and Reid(1987)]{CoxReid87}
Cox, D.R., Reid, N. (1987).
\newblock Parameter orthogonality and approximate conditional inference (with discussion).
\newblock {\em Journal of the Royal Statistical Society B\/} {\bf 40}, 1--39.

\bibitem[D\'iaz--Garc\'ia and Leiva(2005)]{Diaz-Leiva05}
D\'iaz--Garc\'ia, J.A., Leiva, V. (2005).
\newblock A new family of life distributions based on the elliptically contoured distributions.
\newblock {\em Journal of Statistical Planning and Inference\/} {\bf 128}, 445--457.

\bibitem[Doornik(2007)]{DcK2007}
Doornik, J.A. (2007).
\newblock {\em An Object-Oriented Matrix Language -- Ox 5\/}.
\newblock London: Timberlake Consultants Press.
\newblock 5th ed.

\bibitem[Ferrari et al.(1997)]{FerrariBotterCribari1997}
Ferrari, S.L.P., Botter, D.A., Cribari--Neto, F. (1997).
\newblock Local power of three classic criteria in generalised linear models
  with unknown dispersion.
\newblock \textit{Biometrika} \textbf{84}, 482--4S5.

\bibitem[Galea et al.(2004)]{Galea-etal-2004}
Galea, M., Leiva, V., Paula, G.A.~(2004).
Influence diagnostics in log-Birnbaum--Saunders regression models.
\emph{Journal of Applied Statistics} \textbf{31}, 1049--1064.

\bibitem[G\'omes et al.(2009)]{GPB09}
G\'omes, H.W., Olivares--Pacheco, J.F., Bolfarine, H. (2009).
\newblock An extension of the generalized Birnbaum--Saunders distribution.
\newblock {\em Statistics and Probability Letters\/} {\bf 79}, 331--338.

\bibitem[Gradshteyn and Ryzhik(2007)]{GR2007}
Gradshteyn, I.S., Ryzhik, I.M. (2007).
\newblock {\em Table of Integrals, Series, and Products\/}.
\newblock New York: Academic Press.

\bibitem[Guiraud et al.(2009)]{guiraud-et-al-2009}
Guiraud, P., Leiva, V., Fierro, R. (2009).
\newblock A non-central version of the Birnbaum--Saunders distribution for reliability analysis.
\newblock {\em IEEE Transactions on Reliability\/} {\bf 58}, 152--160.

\bibitem[Harris and Peers(1980)]{HarrisPeers1980}
Harris, P., Peers, H.W. (1980).
\newblock The local power of the efficient score test statistic.
\newblock \textit{Biometrika} \textbf{67}, 525--529.

\bibitem[Hayakawa(1975)]{Hayakawa1975}
Hayakawa, T. (1975).
\newblock The likelihood ratio criterion for a composite hypothesis under a
  local alternative.
\newblock \textit{Biometrika} \textbf{62}, 451--460.

\bibitem[Kundu et al.(2008)]{kundu-et-al-2008}
Kundu, D., Kannan, N., Balakrishnan, N. (2008).
\newblock On the function of Birnbaum--Saunders distribution and associated inference.
\newblock {\em Computational Statistics and Data Analysis\/} {\bf 52}, 2692--2702.

\bibitem[Leiva et al.(2007)]{Leiva-etal-2007}
Leiva, V., Barros, M., Paula, G.A., Galea, M.~(2007).
Influence diagnostics in log-Birnbaum--Saunders regression models
with censored data.
\emph{Computational Statistics and Data Analysis} \textbf{51}, 5694--5707.

\bibitem[Leiva et al.(2008)]{Leiva-et-al-2008}
Leiva, V., Barros, M., Paula, G.A., Sanhueza, A. (2008).
Generalized Birnbaum--Saunders distributions applied to air
pollutant concentration. {\em Environmetrics}  {\bf 19}, 235--249 .

\bibitem[Leiva et al.(2009)]{Leiva-et-al-2009}
Leiva, V., Sanhueza, A., Angulo, J.M. (2009).
A length-biased version of the Birnbaum--Saunders distribution with application in water quality.
{\em Stochastic Environmental Research and Risk Assessment}  {\bf 23}, 299--307.

\bibitem[Lemonte et al.(2007)]{LCNV07}
Lemonte, A.J., Cribari--Neto, F., Vasconcellos, K.L.P. (2007).
\newblock Improved statistical inference for the two-parameter
  Birnbaum--Saunders distribution.
\newblock {\em Computational Statistics and Data Analysis\/} {\bf 51}, 4656--4681.

\bibitem[Lemonte and Ferrari(2010a)]{LemFer2010a}
Lemonte, A.J., Ferrari, S.L.P. (2010a).
\newblock The local power of the gradient test.
Working paper {\tt arXiv:1004.5543v1}.

\bibitem[Lemonte and Ferrari(2010b)]{LemFer2010b}
Lemonte, A.J., Ferrari, S.L.P. (2010b).
\newblock Small-sample corrections for score tests in Birnbaum--Saunders regressions.
\newblock {\em Communications in Statistics -- Theory and Methods\/}. {\tt To appear.}

\bibitem[Lemonte et al.(2010)]{Lemonte-et-al-2010}
Lemonte, A.J., Ferrari, S.L.P., Cribari--Neto, F. (2010).
\newblock Improved likelihood inference in Birnbaum--Saunders regressions.
\newblock {\em Computational Statistics and Data Analysis\/} {\bf 54}, 1307--1316.

\bibitem[Lemonte et al.(2008)]{LSCN08}
Lemonte, A.J., Simas, A.B., Cribari--Neto, F. (2008).
\newblock Bootstrap-based improved estimators for the two-parameter
  Birnbaum--Saunders distribution.
\newblock {\em Journal of Statistical Computation and Simulation\/} {\bf 78}, 37--49.

\bibitem[Lepadatu et al.(2005)]{Lepadatu-et-al-2005}
Lepadatu, D., Kobi, A., Hambli, R., Barreau, A. (2005).
\newblock Lifetime multiple response optimization of metal extrusion die.
\newblock {\em Proceedings of the Annual Reliability and Maintainability Symposium\/},
37--42.

\bibitem[Meintanis(2010)]{Meintanis2010}
Meintanis, S.G. (2010). Inference procedures for the Birnbaum--Saunders
distribution and its generalizations. {\em Computational Statistics
and Data Analysis} {\bf 54}, 367--373.

\bibitem[Mittelhammer et al.(2000)]{Mittelhammer-et-al-2000}
Mittelhammer, R.C., Judge, G.G., Miller, D.J. (2000).
\newblock {\em Econometric Foundations}.
\newblock New York: Cambridge University Press.

\bibitem[Rao(2005)]{Rao2005}
Rao, C.R. (2005).
\newblock Score test: historical review and recent developments.
\newblock In {\em Advances in Ranking and Selection, Multiple Comparisons,
and Reliability}, N.~Balakrishnan, N.~Kannan and H.~N. Nagaraja, eds.~Birkhuser, Boston.

\bibitem[Rieck and Nedelman(1991)]{RiekNedelman91}
Rieck, J.R., Nedelman, J.R. (1991).
\newblock A log-linear model for the Birnbaum--Saunders distribution.
\newblock {\em Technometrics\/} {\bf 33}, 51--60.

\bibitem[Saunders(1974)]{Saunders1974}
Saunders, S.C. (1974).
\newblock A family of random variables closed under reciprocation.
\newblock {\em Journal of the American Statistical Association\/} {\bf 69},
  533--539.

\bibitem[Terrell(2002)]{Terrell2002}
Terrell, G.R. (2002).
\newblock The gradient statistic.
\newblock \textit{Computing Science and Statistics} \textbf{34}, 206--215.

\bibitem[Xi and Wei(2007)]{XiWei07}
Xi, F.C., Wei, B.C. (2007).
\newblock Diagnostics analysis for log-Birnbaum--Saunders regression models.
\newblock {\em Computational Statistics and Data Analysis\/} {\bf 51}, 4692--4706.

\bibitem[Xu and Tang(2010)]{XiTang10}
Xu, A., Tang, Y. (2010).
\newblock Reference analysis for Birnbaum--Saunders distribution.
\newblock {\em Computational Statistics and Data Analysis\/} {\bf 54}, 185--192.

\bibitem[Wu and Wong(2004)]{WuWong2004}
Wu, J., Wong, A.C.M. (2004). Improved interval estimation for the
two-parameter Birnbaum--Saunders distribution.
{\em Computational Statistics and Data Analysis\/} {\bf 47}, 809--821.

\end{thebibliography}
}
\end{document}